\documentstyle[aps,twocolumn,prl,graphicx]{revtex}  
\begin{document}
\draft
%
\twocolumn[\hsize\textwidth\columnwidth\hsize\csname 
@twocolumnfalse\endcsname 
%

\title{The energy gap of intermediate-valent SmB$_6$
studied by point-contact  spectroscopy}

\author{K.~Flachbart$^{a,*}$, K.~Gloos$^{b,c}$, E.~Konovalova$^d$,
    Y.~Paderno$^d$, M.~Reiffers$^a$,  P.~Samuely$^a$, P.~\v{S}vec$^e$}

\address{$^a$Institute of Experimental Physics,
  Slovak Academy of Sciences, SK-04353 Ko\v{s}ice, Slovakia}
\address{$^b$Institut    f\"ur   Festk\"orperphysik,
  Technische   Universit\"at  Darmstadt,   D-64289  Darmstadt, Germany}
\address{$^c$Department  of  Physics,  University  of Jyv\"askyl\"a,
  FIN-40351 Jyv\"askyl\"a, Finland}
\address{$^d$Institute  for  Problems  of Materials Science, UA-252680 Kiev, Ukraine}
\address{$^e$Institute of Physics, Slovak  Academy of Sciences,
   SK-84228 Bratislava, Slovakia}

\date{\today}
\maketitle

\begin{abstract}
We have investigated the intermediate valence narrow-gap semiconductor
SmB$_6$  at low temperatures using both  conventional spear-anvil type
point contacts   as  well   as  mechanically   controllable  break junctions.
The zero-bias conductance varied between less than $0.01\,\mu$S and up
to 1\,mS.
The position of the spectral anomalies, which are related to the different
activation energies  and band gaps of SmB$_6$, did not depend on the the
contact size.
Two different regimes of charge transport could be distinguished:
Contacts with large zero - bias conductance are in the diffusive Maxwell
regime. They had spectra with only small non-linearities. Contacts  with
small  zero  -  bias  conductance are in the tunnelling regime. They had larger
anomalies, but still indicating a finite $45\%$ residual quasiparticle density
of states at the Fermi level at low temperatures of $T=0.1\,$K.
The density of states derived from the tunelling spectra can be decomposed
into two energy-dependent parts with $E_g=21\,$meV and $E_d=4.5\,$meV wide
gaps, respectively.
\end{abstract}



%
] 
\narrowtext
%

\section{Introduction}
Samarium hexaboride SmB$_{6}$  is a homogeneous intermediate
- valence compound in which   the electronic structure at
low temperatures shows  a narrow energy gap as  well as spin
gap,  both originating  from the  hybridization between  the
narrow states  formed by electrons of  samarium 4$f$ - shell
and  the wide  conduction band  formed by  both boron  $p$ -
states and  Sm 6$s$ -  states \cite{Wachter94}. A  review of
this   and  similar   materials  can   be  found   in  Refs.
\cite{Wachter94,Aeppli92,Kasuya94,Riseborough00}.

Recent experiments on SmB$_{6}$ \cite{Batko93,Nanba93,%
Cooley95a,Cooley95b,Nyhus97,Roman97,Sluchanko99,Gorshunov99,Dressel99}
have shown that in the low-energy  excitation spectrum of  this material
several energy   scales   exist,   and   also   several  regimes  of
low-temperature electron kinetics.  At least three different
activation   energies   determine   the   behaviour  of  the
conduction   electrons.   In   the  temperature  range
$70\,{\rm{K}} >  T > 15\,$K the  properties of SmB$_{6}$ are
governed  by  the  hybridization   gap  $E_g  \approx  10  -
20\,$meV.
Between $15\,$K and $5\,$K, a narrow in - gap band separated
from  the bottom  of the  conduction  band by a direct
activation  energy  of  $E_d  \approx  3  - 5\,$meV has been
observed. The properties of this  narrow band seem to  be
influenced by the  content  of  impurities  and  imperfections  of the
specific sample.
Below  about  $5\,$K  the  electrical conductivity saturates, indicating a
small conductivity  channel within the $E_d$ in - gap states, where the Fermi
level is pinned.

Various models  have been proposed to  explain the formation of  the
$E_d$  -  band  and  the  origin  of  the  residual conductivity
\cite{Kasuya94,Allen79,Mott82,Kasuya79,Kikoin95,Curnoe00,Hanzawa98,%
Kebede96,Park98}, but so far no final conclusion could be obtained.
While Ref.~\cite{Gorshunov99} favours hopping processes of electrons,
Refs.~\cite{Sluchanko00,Gabani01} prefer a coherent metal-like state at
low temperatures.
In either case electrons are strongly localized ($\sim 0.6\,$nm localisation
radius) at random impurities with a small concentration of $N \approx
10^{23}\,{\rm{m}}^{-3}$ at low temperatures \cite{Sluchanko00}.

To search for anomalies in the quasi-particle density of states a number of
experiments on small junctions with SmB$_6$ has been performed.
Frankowski and Wachter \cite{Frankowski82} brought a sharply etched
molybdanum tip into direct contact with the surface of a cleaved  SmB$_6$
single crystal.
The spectra of these low-resistance contacts showed a 4.6\,meV wide anomaly
(full width at half maximum) which was attributed to a gap in the density of
states. The overall size of this anomaly was only about $10\%$ of the mean
contact resistance.
G\"untherodt  {\em et al.} \cite{Guentherodt82} investigated Schottky-type
tunnel contacts between SmB$_6$ and GaAs. They found a huge but only
2.7\,meV wide zero-bias anomaly. Such small gaps were obtained only when
the SmB$_6$ surface was sputter cleaned {\em in situ}. Without such a
treatment the anomalies broadened to around 10\,meV.
Planar tunnel junctions with lead counter electrodes were investigated
by Batlogg {\em et al.} \cite{Batlogg81} and  by Amsler {\em et al.}
\cite{Amsler98}. The latter experiments showed 14\,meV wide spectral
anomalies, roughly coinciding with the $E_g$ band gap,  and a $\sim 70\%$
residual quasiparticle density of states at the Fermi level.
No consistent picture could evolve from all these experiments, because the
width as well as the size of the main zero-bias anomaly varied a lot.
To obtain more and, hopefully,  reliable information on the
different energy scales involved we investigated direct junctions between
two bulk pieces of SmB$_6$ as function of contact size.

\section{Experimental}
Our SmB$_6$  samples were  cut from  one batch which  was  grown
by  the  zone-floating  method  as described in Ref.~\cite{Konovalova90}.
About   800   ppm   magnetic   impurities,  mostly  magnetic
lanthanide elements close to Sm  in the periodic table, were
found by magnetic susceptibility measurements \cite{Roman97}.
A similar   amount   of   non-magnetic   impurities,  mainly
lanthanum,   was   detected   by   induction-coupled  plasma
spectroscopy.
sufficient size.

Two different type of point contacts were prepared.
First, two mechanically polished  pieces of SmB$_{6}$ were brought into
contact in a spear-anvil type setup and measured at $4.2\,$K  in liquid helium.
Second, bulk pieces of SmB$_{6}$ were broken at a predefined notch  in the
ultrahigh vacuum  region of  a cold  $^3$He-$^4$He  dilution refrigerator and
measured  mainly at 0.1\,K. Reference \cite{Gloos99} describes our
break-junction apparatus in detail. With mechanically-controllable break
junctions  we avoid the  oxidation of  the interfaces.
It also offers exellent mechanical stability of the contacts so that also
very small junctions can be investigated.
In both  cases the $dI/dU(U)$  spectra were recorded
in the  standard four - terminal  mode with current biasing.

\section{Electrical  conductivity and activation energies}
\begin{figure}
\centerline{\includegraphics[width=7cm]{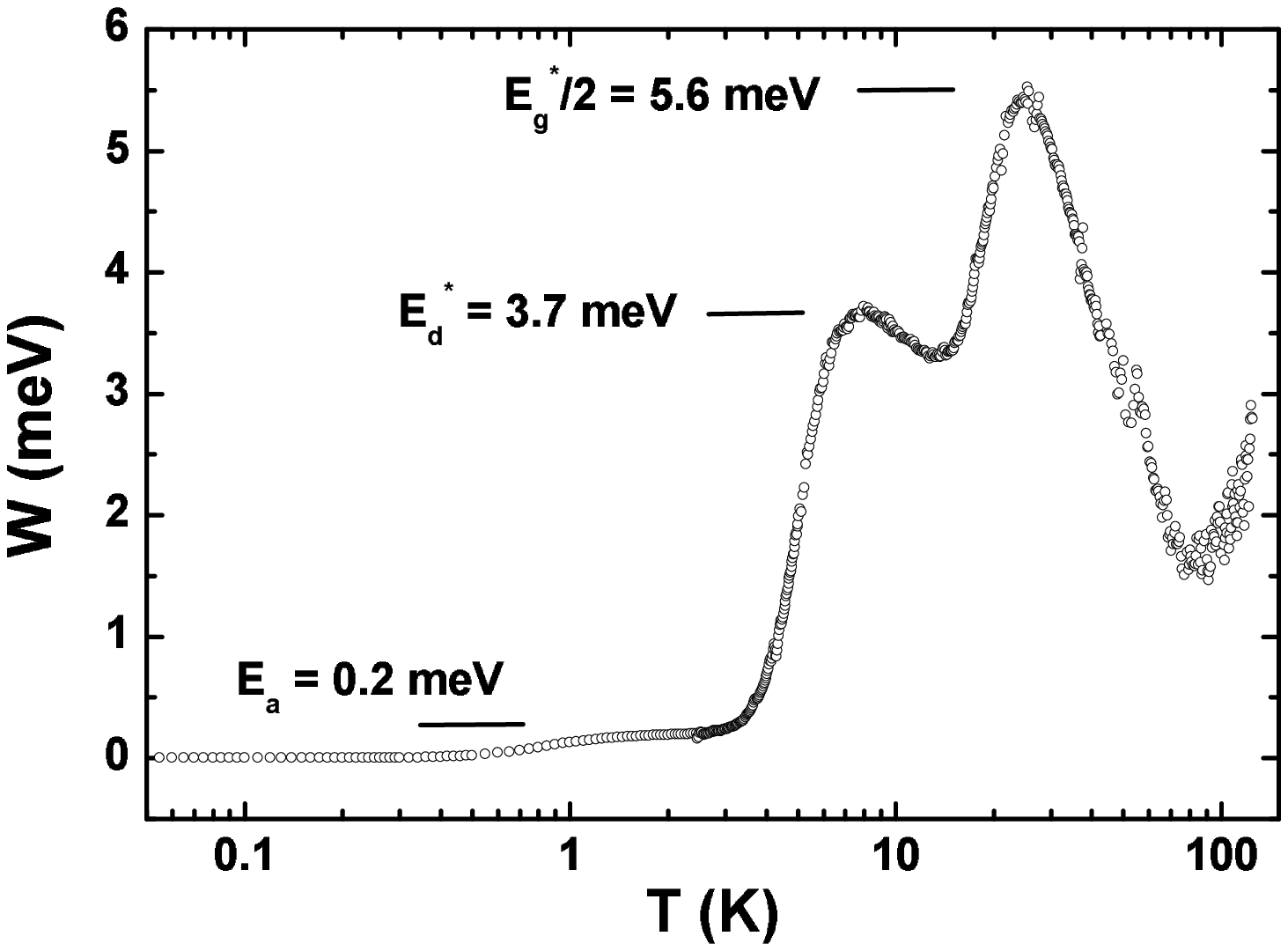}}
\caption{Temperature  dependence of the  activation energy $W$ of  SmB$_6$,
calculated  from   the  conductivity  data  in Fig.~\ref{fig-activate}.}
\label{fig-activate}                        
\end{figure}

Figure \ref{fig-sigma}  shows  the  temperature  dependence  of  the  bulk
conductivity   $\sigma$($T$)  of   the  single   crystalline SmB$_{6}$ sample.
$\sigma$$(T$)  decreases in a rather complicated  way with  decreasing $T$,
saturating at the lowest temperatures. Describing the conductivity by
thermally activated mechanism, $\sigma(T)
\propto \exp{(-W/k_{\rm{B}}}T)$, an activation energy
\begin{equation}
  W = -d(\ln{\sigma(T)} / d(1/k_{\rm{B}}T)
  \end{equation}
can be defined (Fig.~\ref{fig-activate}). It shows a peak of about 5.6\,meV
in the temperature range between 70\,K and 15\,K.
This  value can be attributed to $E_g^*/2$ when the Fermi level sits just in
the center of the hybridization band gap of width $E_g^*  \approx 11.2\,$meV.
Between  15\,K and 5\,K the second pronounced peak apppears.
Its value of $E_d^* \approx 3.7\,$meV corresponds to the formation of the
narrow in - gap band within the hybridization gap located $E_d^*$ below
the conduction band edge.
These two activation energies do not clearly level off.
Therefore the index '*' is used to distinguish the activation-derived values
from the true energy gaps still to be determined.

With further decreasing temperature, the activation energy $W$ falls to very
low  values.  The anomaly  at about  $1-2\,$K with  $E_a
\approx  0.2\,$meV  originates  probably  from an additional
narrow band formed inside the energy gap (close to the $E_d$
- band) due to the relatively large content of impurities of
the sample.
Below  about 1\,K,  however, the  activation energy is lower  than the
available  thermal energy $k_{\rm{B}}T$, indicating  a  metallic - like
conduction  mechanism at the lowest  temperatures.
This would agree with recent $ac$ -  conductivity \cite{Sluchanko00} and
specific  heat measurements of SmB$_{6}$  \cite{Gabani01},  which  are
interpreted as showing a  transition  into  a coherent  heavy-fermion like state
below about 5\,K.

\section{Point-contact spectroscopy}
\begin{figure}
\centerline{\includegraphics[width=7cm]{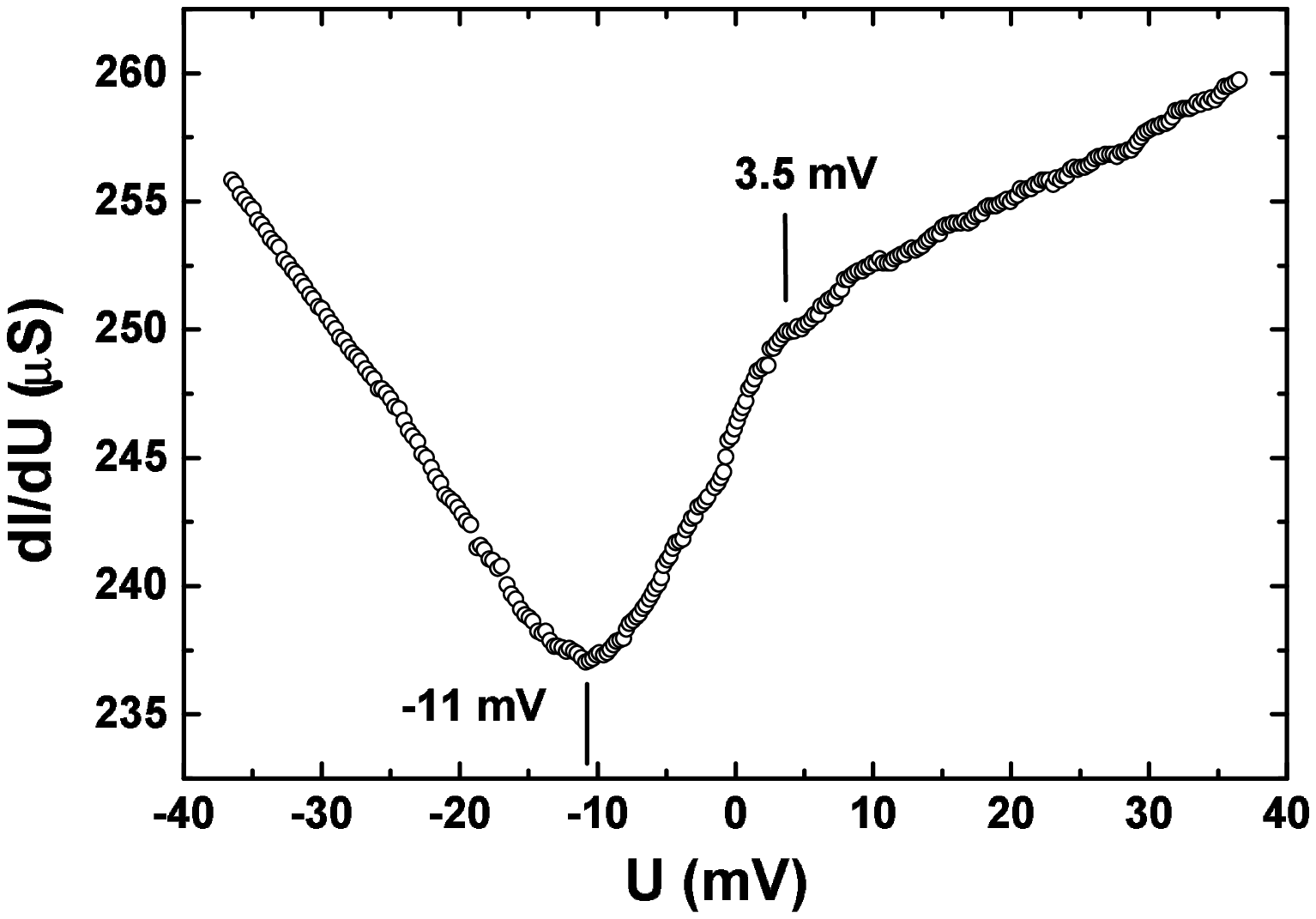}}
\caption{Symmetric (top) and asymmetric (bottom) $dI/dU(U)$ spectra of
spear-anvil type contacts with high zero-bias conductance. $T = 4.2\,$K.}
\label{fig-spear}                        
\end{figure}

We have investigated more than 100 SmB$_{6}$ - SmB$_{6}$ point contacts with
zero-bias conductance ranging  from about 0.01~$\mu$S to about 1\,mS, most of
them using the spear-anvil technique. These  spear-anvil type junctions had a
high zero - bias conductance and usually symmetric $dI/dU(U)$ spectra,
similar to those  obtained earlier  on  SmB$_{6}$  - Mo  point contacts
\cite{Frankowski82}.
But few of those junctions were asymmetric like that in Figure \ref{fig-spear}.
We found anomalies, typically small kinks in the $dI/dU$ spectra indicating a
change of slope, at $U \approx 4\,$mV and $U \approx 12\,$mV, respectively.

\begin{figure}
\centerline{\includegraphics[width=7cm]{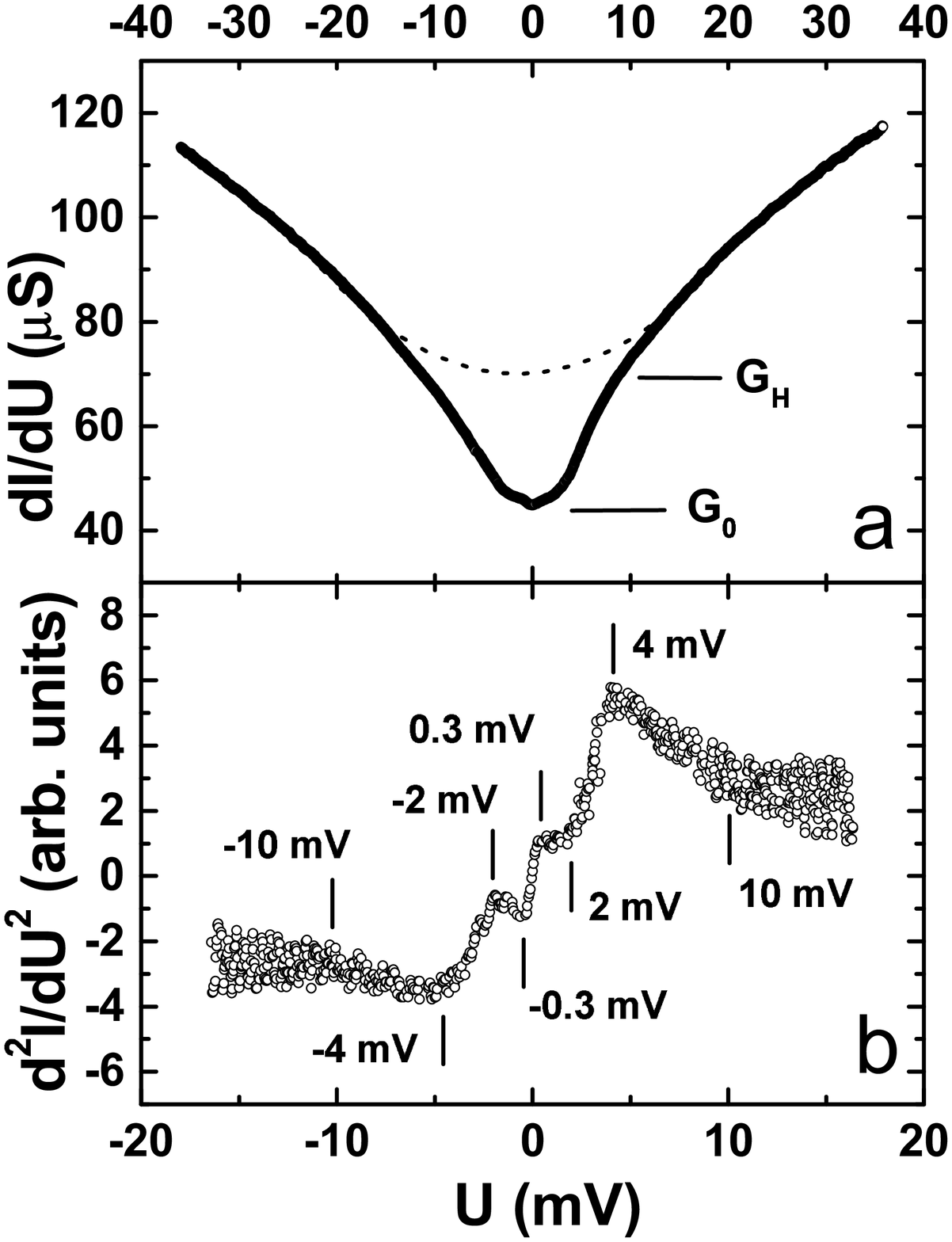}}
\caption{(a) $dI/dU(U)$ spectrum of a contact with high zero-bias conductance
$G_0=dI/dU(U=0)$ at $T  =  0.1\,$K. The dotted line describes tentatively the
expected spectrum at high temperatures with zero-bias conductance $G_H$.
(b) Second derivative $d^2I/dU^2(U)$ of the spectrum in Figure (a).}
\label{fig-hi-G}                        
\end{figure}

\begin{figure}
\centerline{\includegraphics[width=7cm]{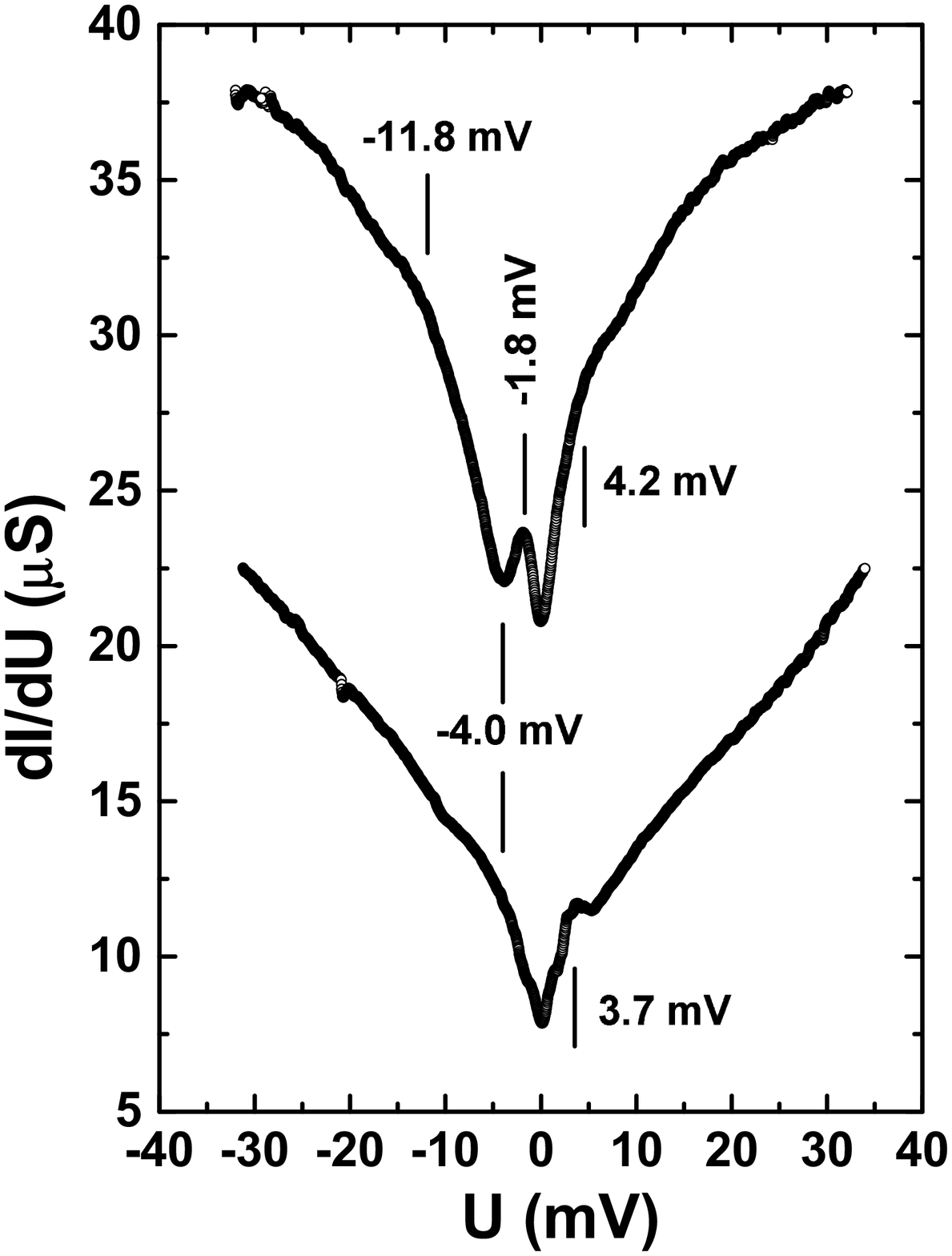}}
\caption{$dI/dU(U)$ spectra of contacts  with medium zero - bias conductance
at $T = 0.1\,$K. Arrows mark the  characteristic energies.}
\label{fig-med-G}                        
\end{figure}

\begin{figure}
\centerline{\includegraphics[width=7cm]{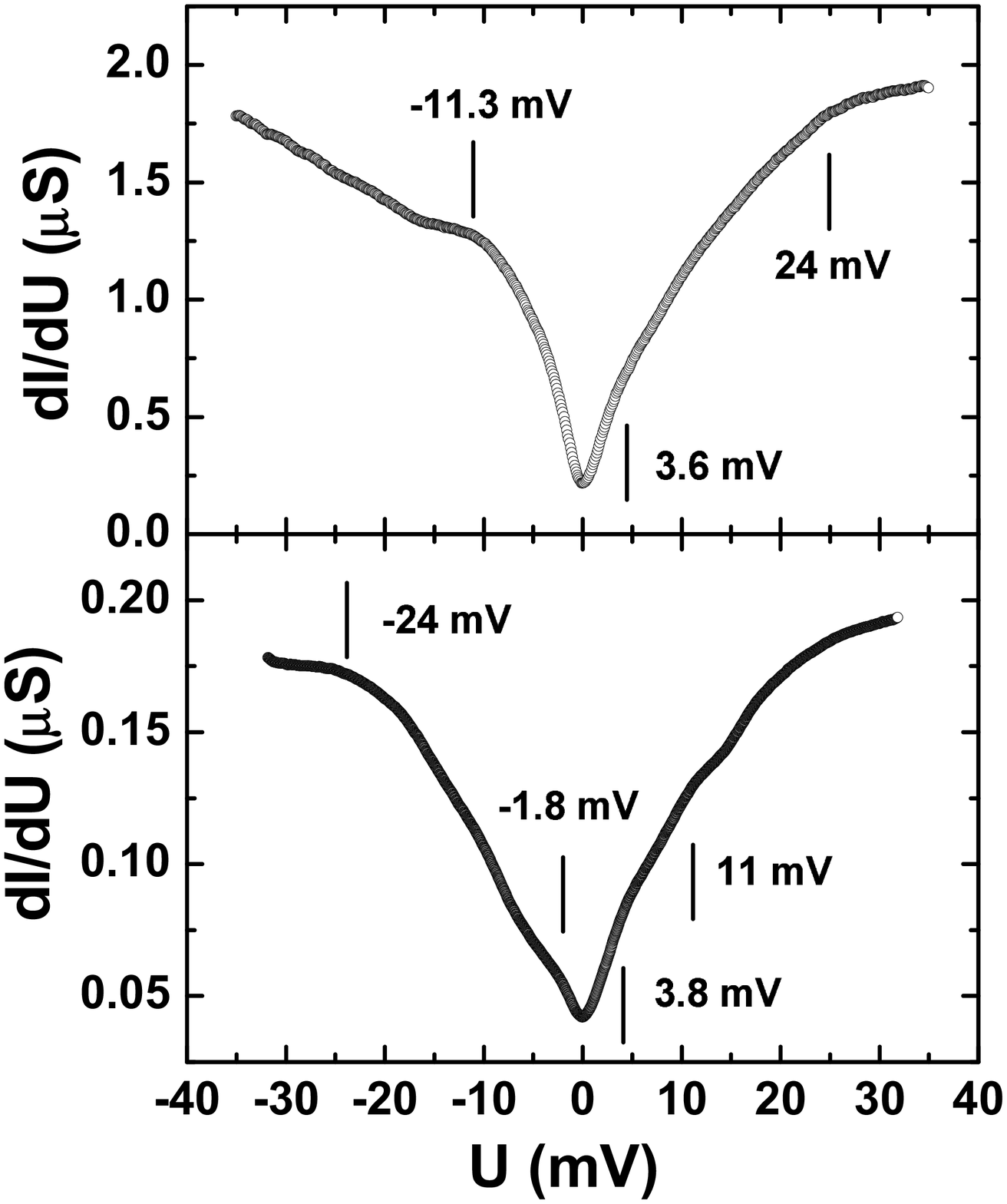}}
\caption{$dI/dU(U)$ spectra of contacts  with low zero - bias conductance at
$T = 0.1\,$K. Arrows mark the  characteristic energies.}
\label{fig-lo-G}                        
\end{figure}

Figures \ref{fig-hi-G}, \ref{fig-med-G}, and \ref{fig-lo-G} show typical $dI/dU$
spectra of SmB$_{6}$ break junctions at $T=0.1\,$K.
Like the spear-anvil type contacts they had anomalies in $dI/dU$  and in
$d^2I/dU^2$ at about 4\,mV and 12\,meV.
Very often additional anomalies were observed close to 0.2\,mV, 1.8\,mV, and
22\,meV. Several junctions also had anomalies at  about  80\,mV.

Spectra with high  zero-bias  conductance  $G_0 = dI/dU(T=0.1\,K, U=0)$
shown in Fig.~\ref{fig-hi-G} were symmetric with respect to the applied
voltage. Figs.~\ref{fig-med-G} and \ref{fig-lo-G} show that when the zero-bias
conductance $G_0$ falls below about $30\,\mu$S, the differential
conductance becomes more and more asymmetric.
This tendency towards asymmetry could result from an inhomogeneous
distribution   of   impurities  and imperfections in  the sample which can
lead  to different local carrier concentration, and which become the
more pronounced the smaller the contact interface is.
The different possible crystallographic orientations of the SmB$_6$ electrodes
forming the  contact could also contribute to the asymmetry.

The spectra of junctions with low conductance (Fig.~\ref{fig-lo-G}) appear to
be broader than those with higher zero-bias conductance (Fig.~\ref{fig-med-G}).
However, according to Fig.~\ref{fig-pos} the position of the anomalies
is not affected by any variation of $G_0$. This points then to different
weighing factors of the anomalies at small and at large zero-bias conductance,
respectively,  possibly reflecting the two different regimes of
charge transport discussed below.
In Fig.~\ref{fig-pos} we have labeled the positions according to the activation
energies $E_a$, $E_d^*$, and $E_g^*$.
Anomalies around 1.8\,mV as well as around 22\,meV could be related with
$E_g^*/2 -  E_d^*$ and $2E_g^*$, respectively.
And one could speculate whether the 80\,mV anomalies (not shown in the figure)
correspond  to the excitation of electrons into higher  energy levels of the
Sm - ions, considered   for example to interprete magnetic properties of
SmB$_6$ \cite{Wachter94,Nickerson71}.

\begin{figure}
\centerline{\includegraphics[width=7cm]{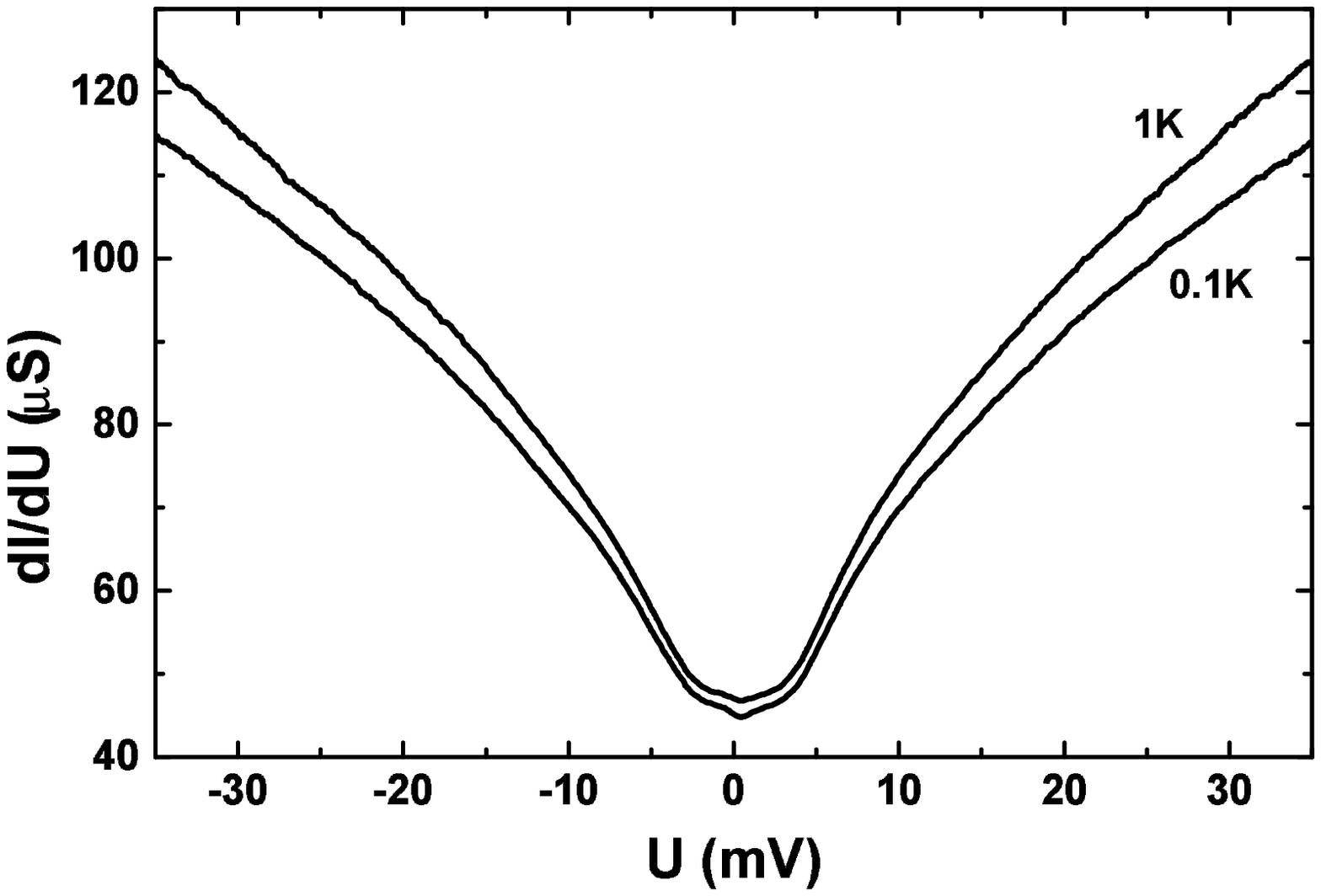}}
\caption{$dI/dU(U)$ spectra of  a contact at $T=0.1\,$K and $T=1\,$K,
respectively.}
\label{fig-temp}                        
\end{figure}

\begin{figure}
\centerline{\includegraphics[width=7cm]{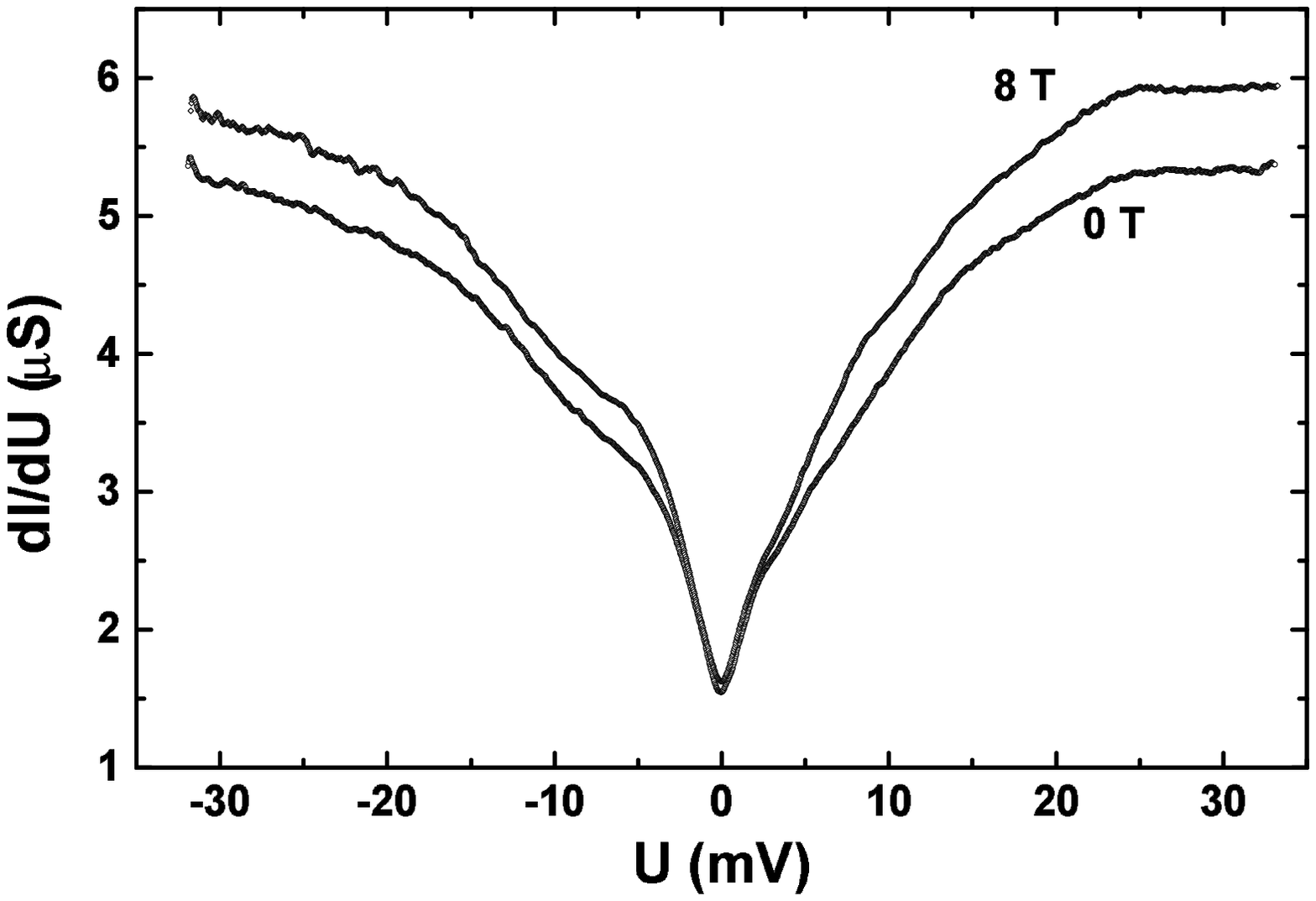}}
\caption{$dI/dU(U)$ spectra of  a contact at $T = 0.1\,$K in  a magnetic
field of $B=0$ and $B=8\,$T, respectively.}
\label{fig-bf}                              
\end{figure}

\begin{figure}
\centerline{\includegraphics[width=7cm]{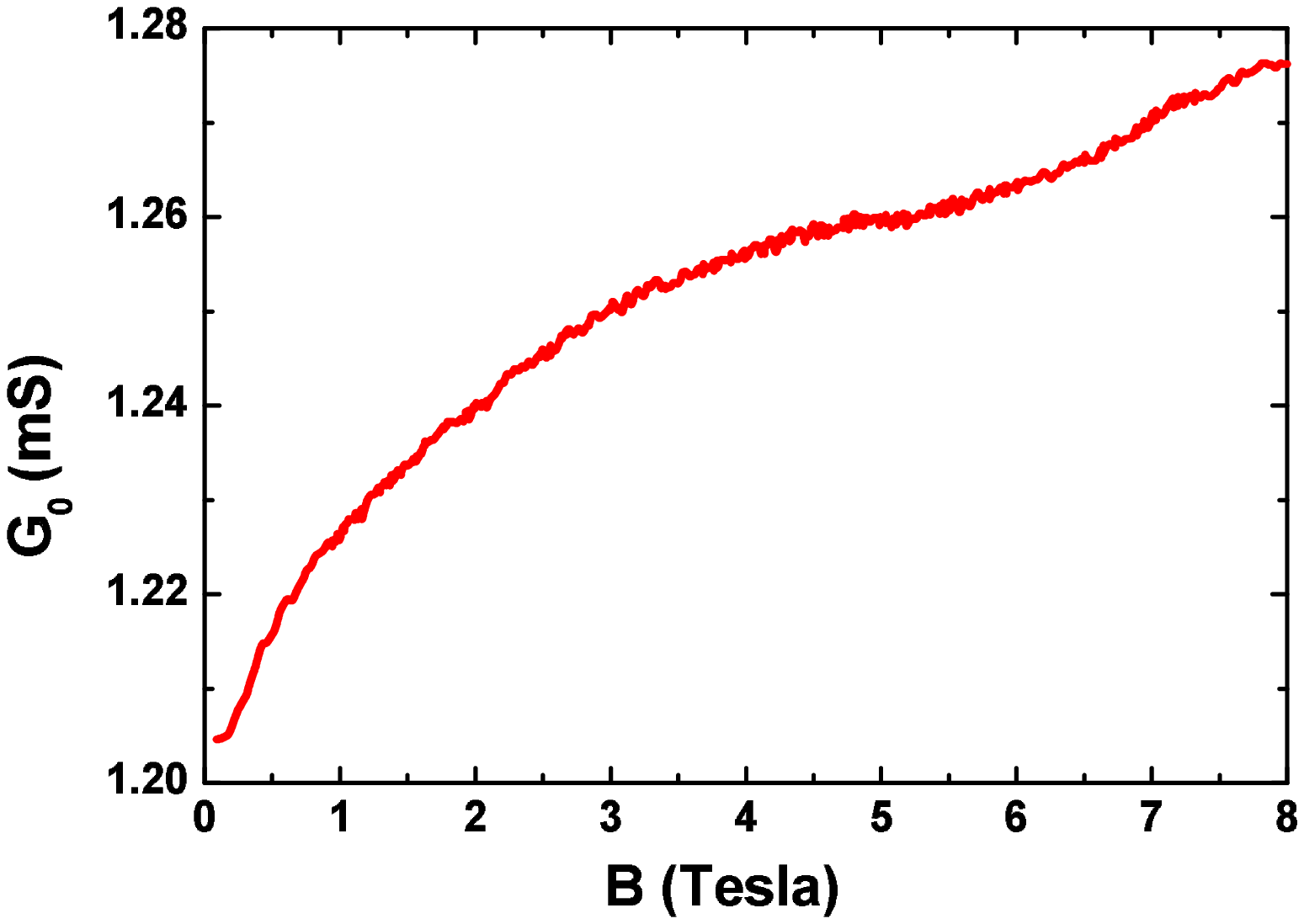}}
\caption{$G_0(B)$ of one contact at $T = 0.1\,$K.}
\label{fig-magres}                              
\end{figure}

The properties of the junctions depend only weakly on temperature
(Fig.~\ref{fig-temp}) and on magnetic field (Fig.~\ref{fig-bf}). In both cases,
the shape of the spectra does not change much, only the conductance
increases slightly when either the temperature is raised from
0.1\,K to 1\,K or the magnetic field from 0\,Tesla to 8\,Tesla (the field was
always perpendicular to the direction of current flow).
Thus also the magneto resistance is small and negative like in the bulk sample.
Junctions with large zero-bias conductance $G_0 \approx 1\,$mS had typically
$G_0(B=0) / G_0(B) - 1 \approx -1.3\times 10^{-3} B^2$,  where $B$ is the
applied field in Tesla (Fig.~\ref{fig-magres}).

\section{Regimes of charge transport}
To  evaluate   the  observed  $dI/dU(U)$ dependencies not only qualitatively,
one should know the regime of charge transport across the contacts.
This is absolutely necessary if one wants to attribute a spectral
anomaly at a bias voltage $U$ to an anomaly in the density of states
or to some scattering mechanism at an energy $eU$.
We face here a quite general -- but very often ignored -- problem of direct
contacts (that means contacts without well-defined tunneling barrier) between
conductors with a short electronic mean free path.
Naidyuk and Yanson \cite{Naidyuk98} have recently reviewed and dicussed
this topic.

As one typical example to illustrate the situation we
refer to point-contact experiments  with the Kondo semiconductor CeNiSn.
It is ususally believed that direct junctions with this short mean free path
compound are in the tunnelling regime, implying that the spectra measure
the density of states \cite{Ekino95,Davydov97}.
However, it was demonstrated recently that those point contacts are more
likely metallic \cite{Naidyuk00}.
And this, in turn, makes it impossible to extract the density of states.

We emphasize that the shape of the spectra itself does not  tell anything
about the regime of charge transport, whether it is tunnelling or diffusive
transport, for example, unless one works with a well-known substance
for which the spectra can be predicted reliably.
Experiments with a superconducting lead counter electrode like that in
Ref.~\cite{Amsler98} allow the quality of a tunnel junction to be verified.
For such an experiment quantum tunneling can be confirmed or disproved.
Obviously, this does not work for SmB$_6$ in contact with an other piece of
SmB$_6$, since the exact properties of this compound have yet to be
determined.
Thus for our experiments the situation seems hopeless. But there exists
what we believe are strong arguments  to identify the relevant  transport
mechanisms. We use a procedure that has been developed for contacts with
heavy-fermion compounds \cite{Gloos96,Gloos98} as well as for junctions with
doped germanium \cite{Sandow00,Sandow01}, both materials with a typically
short electronic mean free path (or hopping length in the latter case).

\section{Large junctions -- Maxwell regime}
\begin{figure}
\centerline{\includegraphics[width=7cm]{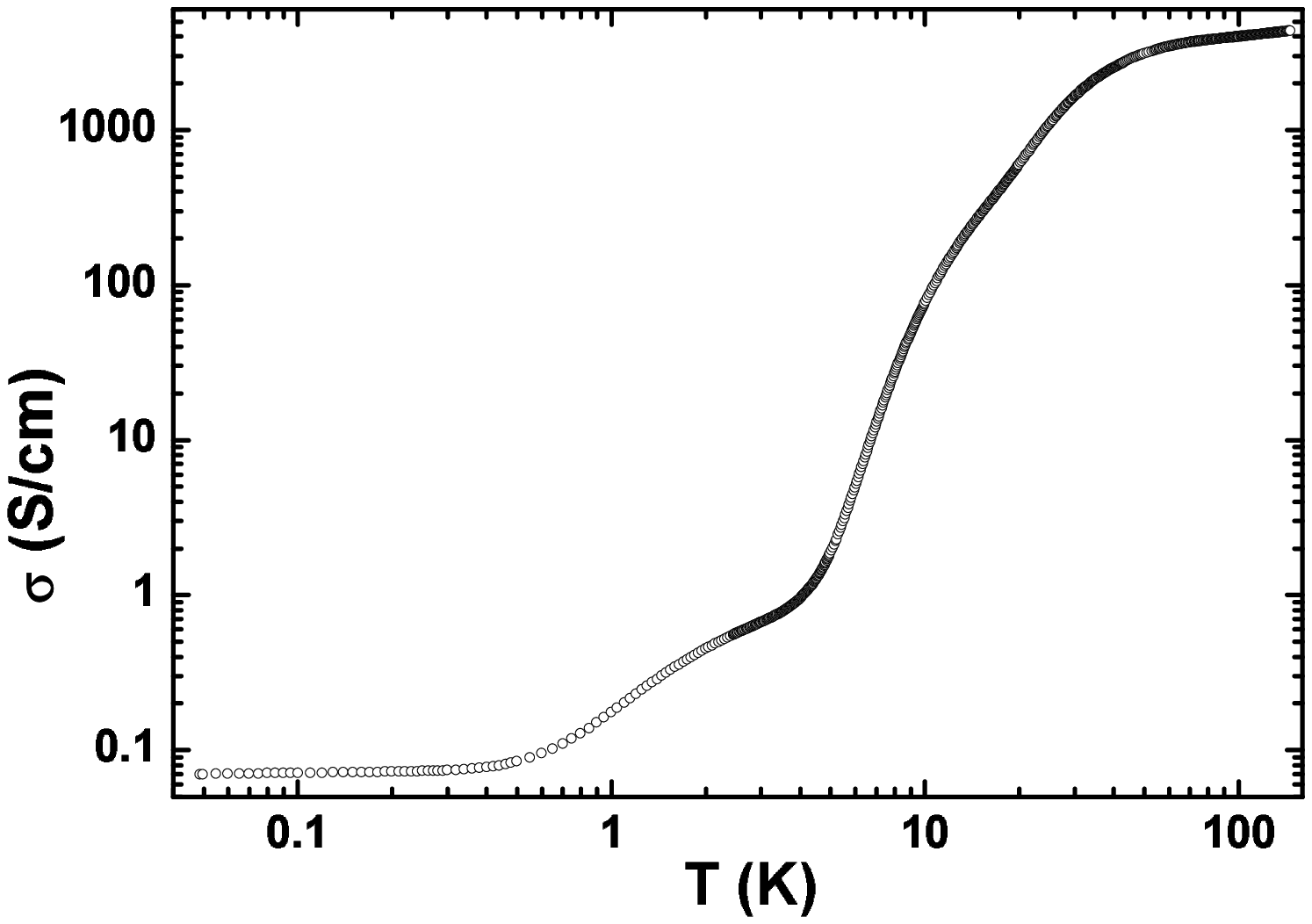}}
\caption{Electrical conductivity $\sigma$ vs.~ temperature $T$ of the bulk
SmB$_6$ sample.}
\label{fig-sigma}                  
\end{figure}

SmB$_6$ has -- due  to  hybridization  effects --  an extremely short
electronic mean free path at low temperatures, which amounts to not
more than several lattice constants.
Therefore it is difficult to evaluate our experiments  in
terms  of  the classical metallic point contact spectroscopy.
When the diameter $d$, the characteristic dimension of the point-contact,
is large, it can certainly be estimated using Maxwell's formula
$G \approx \sigma d$, where $G = dI/dU$ is the point-contact conductance
at zero bias (a circular shape of the contact is a useful simplifying assumption).
Two approaches are possible:
One could set $G$ of Maxwell's formula equal to the measured $dI/dU(U=0)$
and use the measured $\sigma$ at the same temperature. Thus, with
$\sigma (T=0.1\,{\rm{K}}) \approx 0.05\,\mu$S/cm from Fig.~\ref{fig-sigma}, a
$G_0  = 1\,\mu$S junction has a $d \approx 200\,$nm diameter.
However, it is better -- and this has been clearly demonstrated for junctions
with the heavy-fermion superconductors \cite{Gloos96,Gloos98} -- to use the
change of conductance $\delta G$ and conductivity $\delta \sigma$,
respectively, when the temperature is increased.
Thus the contact diameter in the Maxwell regime actually becomes
\begin{equation}
  d \approx \delta \sigma / \delta G .
  \label{Maxwell-equation}
  \end{equation}
We have noted that the conductance increases by roughly  $5\,\%$ when the
temperature is increased from 0.1 to 1\,K.
At the same time, the bulk conductivity of our sample increases by a
factor of 4 to about 0.2\,S/cm.
If we attribute this increase of conductance to the increase of bulk
conductivity in Fig.~\ref{fig-sigma} then, according to
Eq.~\ref{Maxwell-equation}, the diameter amounts to about 3\,nm only.
(Applying this method on SmB$_6$ is by far less reliable than for junctions
with the heavy-fermion compounds. The above value should be regarded
as an estimate for the order of magnitude of $d$.)
Such a small contact is still formed by $\sim 100$ atoms which contribute to
its mechanical stability. In fact, we have found a lack of stability, indicating
the transition to 'one atom' junctions, only when the conductance is further
reduced by a factor of at least 100.
This qualitative agreement supports our estimate of the contact size.
Nevertheless, Maxwell's formula is valid only for large junctions, and we have
yet to determine its lower bound.

The large temperature-independent background needs to be explained:
It is not due to impurities of the   bulk sample.
More likely the contact distorts the crystal lattice locally, and these
additional defects enhance the local conductivity.
A similar phenomenom -- but of opposite direction -- has been observed at
junctions with heavy-fermion materials \cite{Gloos96,Gloos98}.
Lattice defects there enhance the local resistivity and, consequently,
the contact resistance.
Using the absolute value of the contact resistance itself instead of its
temperature-dependent part would then considerably underestimate the contact
diameter, contrary to the SmB$_6$ junctions.

Joule heating of our SmB$_6$ - SmB$_6$ contacts could be expected at high
bias voltages. A simple heating model \cite{Kulik92} predicts the effective
temperature of the contact
\begin{equation}
  T_{eff} = \sqrt{T_b^{2} + U^2/4L}
  \label{joule-heating}
  \end{equation}
at a bath temperature $T_b$. Here the Lorenz number $L  =  \kappa/\sigma T$,
and $\kappa$ denotes  the thermal conductivity of  the material in the
contact region.
In the Drude  - Sommerfeld theory of metals $L$ equals $L_0 = 2.45
\times 10^{-8}\,$V$^{2}/$K$^{2}$, implying a significant overheating of
the  contact.
However,   SmB$_{6}$  has a  very low   electrical
conductivity  below about  5~K  (Fig.~\ref{fig-sigma})  but a  rather high
thermal conductivity \cite{Flachbart82} due to heat transport by phonons.
This  yields a  more than five orders of magnitude higher $L$ and
provides a very effective cooling of the contact.
Electrons crossing such a large junction take part in many scattering events,
but basically they keep their original kinetic energy as they diffuse through the
contact region.
At a bath temperature of  $T_b  = 0.1\,$K a voltage of $U = 30\,$mV would heat
up the contact to not more than $T_{eff}  \approx 0.4\,$K.
Therefore  the  applied voltage  $U$   is too small to account for any of the
observed anomalies of the $dI/dU$ spectra in terms  of overheating, and
the observed anomalies should mark the correct energies at which
certain energy bands can be occupied or scattering mechanisms start to work.

The  small magnetoresistance, less than $10\%$ at $B = 8\,$Tesla for the
junction in Fig.~\ref{fig-bf}, also indicates  no overheating. Otherwise a
much larger effect had to be expected from the strong magnetoresistivity
of the bulk sample above 1\,K.

\section{Small junctions -- Tunnelling}
On further reducing the lateral size or diameter of the junctions one should
expect a transition to ballistic transport  if SmB$_6$ was an ordinary metal.
Obviously, this is not the case. Because of the low carrier concentration of
$N \approx 10^{23}\,{\rm{m}}^{-3}$ \cite{Sluchanko00}, the junctions will
instead undergo  a  transition to tunnelling.
Again there are two different approaches to estimate the contact diameter
at which this transition will take place, depending on whether conduction
is due to electrons in a coherent metal-like state or due to hopping.

First, if the electron system at low temperatures is described by a coherent
state, the Fermi wave number amounts to about
$k_{\rm{F}}  = (3\pi N)^{1/3} =   10^8\,{\rm{m}}^{-1}$.
In the contact region, discrete energy levels can exist which carry
electrical  current as long as the contact diameter $d$ is larger than
$4/k_{\rm{F}} \approx 40\,$nm. If the diameter is smaller, electrons have to
cross the junction as evanescent waves, that means they tunnel.
Such a situation has been observed at direct junctions with antimon, also a
compound with a low carrier density \cite{Krans94}.

Second, if electron transport is due to thermally activated hopping between
localized impurity states as proposed in Ref.\cite{Gorshunov99}, for example,
then junctions with SmB$_6$ should be compared with those of a real
semiconductor like germanium.
Junctions with doped  Ge seem to be in a tunnelling state  although there is
a direct contact with the bulk material without dielectric barrier
\cite{Sandow00,Sandow01}.
Transport across such contacts can be called 'tunnelling' because
the hopping length is so large that the charge carriers cross the junction
interface by one single hop process. These hop processes across the
junction are the same as that in the bulk material.
We believe that our SmB$_6$ junctions with small zero-bias conductance
can be in such a tunneling-like state.
A reasonable estimate for the impurity density is the carrier density, and
the average distance betwen the impurities equals the hopping length.
The transition to tunneling has then to be expected when the contacts become
smaller than $N^{-1/3} \approx 22\,$nm.

For both approaches, using Eq.~\ref{Maxwell-equation} to estimate the
contact diameter, the transition towards tunnelling should take place when
the zero-bias conductance gets smaller than about $10\,\mu$S.
And this is just what we are observing: a transition in the size of the
zero-bias anomalies when $G_0$ is reduced.
We fit a parabola to the spectra a high voltages as indicated in
Fig.~\ref{fig-hi-G} to define as reference the conductance at zero bias $G_H$
if there was no gap in the density o states.
(According to Ref.~\cite{Amsler98} one would have to increase the temperature
of a specific junction to above 40\,K to get an experimental value for $G_H$.
This was not possible with our apparatus). The relative size of the
zero-bias anomaly at low temperatures is  then $G_0/G_H$.
For contacts between SmB$_6$ and a normal metal with constant density
of states around the Fermi level, like that in Ref.~\cite{Amsler98},
the differential conductance $dI/dU(U)$ is
proportional to the energy dependence of the local electronic density of states
multiplied  by the transmission probability of the contact  \cite{Padovani71}.
Thus  $g_0/g_H \approx  G_0/G_H$ when $g_0$, the low-temperature density of
states at the Fermi level, is normalized to the density of states without gap
$g_H$. Since  for SmB$_6$ - SmB$_6$ homocontacts the same quasiparticle
density of states on both sides of the contact contribute to the tunnel
conductance, $g_0$ has to be described by
\begin{equation}
  g_0/g_H \approx  \sqrt{G_0/G_H}
  \label{qp-dos}
  \end{equation}
Fig.~\ref{fig-dos} shows that large junctions (large $G_0$) also have large
$\sqrt{G_0/G_H}$ because of their large background signal. This is
inconsist with the quasiparticle density of states of around $70\%$ measured
using planar tunnel junctions at $T=10\,$K \cite{Amsler98}. Moreover there
seems to be a systematic decrease of the relative size of the anomaly as $G_0
\rightarrow 1\,$mS, possibly indicating a transition from diffusive to thermal
transport when progressively more scattering processes are required for an
electron to pass the contact region. Its initial kinetic energy is then not
conserved any more.
The signal size of smaller junctions, however,  agrees much better with the
expected density of states if we take into account our much lower temperatures
of $T=0.1\,$K, letting us attribute the zero-bias conductance
to the finite density of quasi-particles in the gap at the Fermi level.
Thus bulk transport turns to tunneling below about $G_0 \approx 2\,\mu$S,
in excellent agreement with our above estimate.

\section{The energy gaps and the quasiparticle density of states}
\begin{figure}
\centerline{\includegraphics[width=7cm]{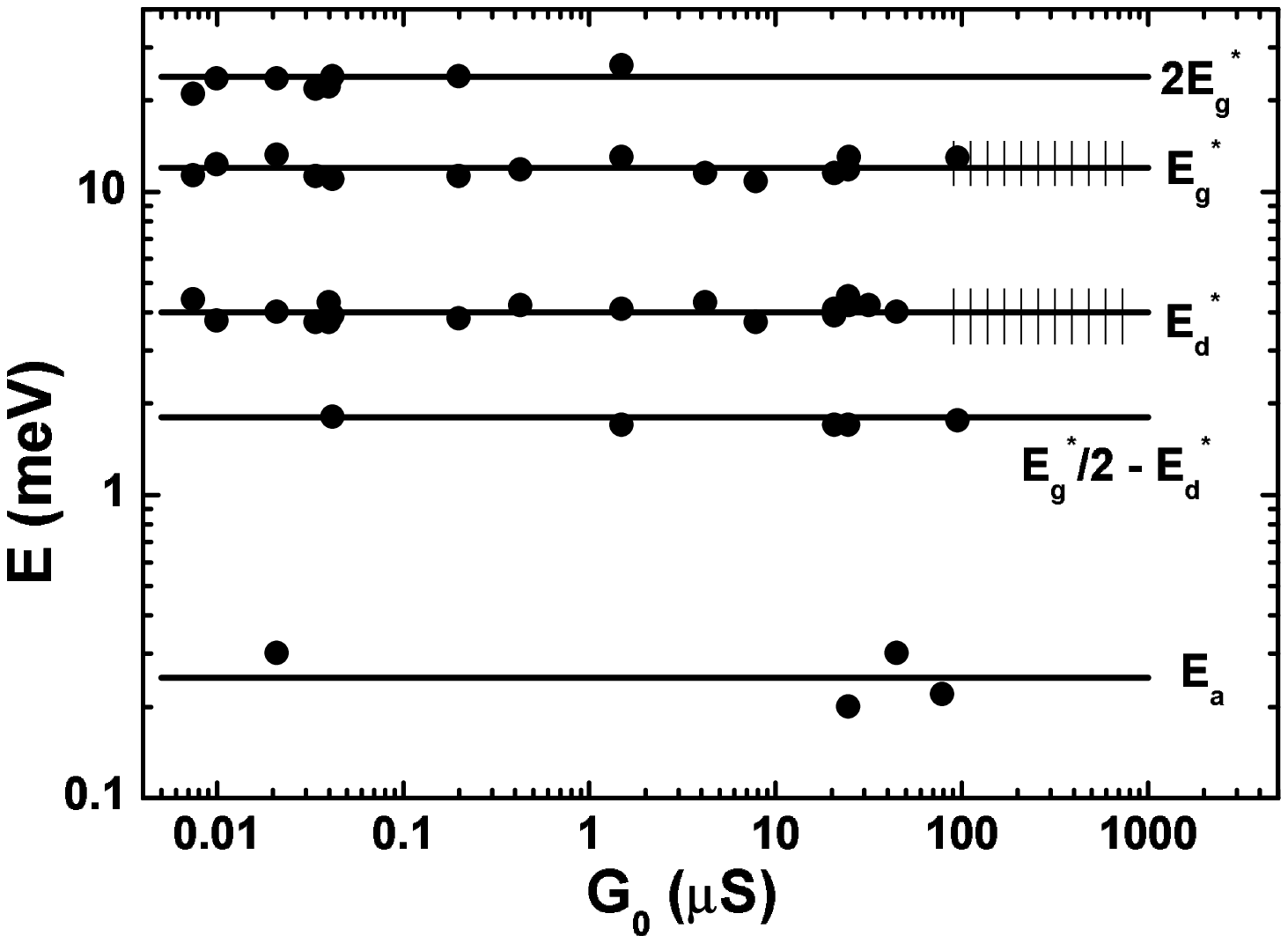}}
\caption{Position of the anomalies vs.~conductance at zero bias voltage.
Solid circles represent break-junction data, the shaded area indicates average
values derived from the spear-anvil type junctions (for which $G_0$ is
the zero-bias conductance at 4.2\,K).
Solid lines are guides to the eye, and tentatively labeled with the
corresponding activation energies.}
\label{fig-pos}                              
\end{figure}

Most of the anomalies, for the break junctions as well as for the spear-anvil
type contacts, were found at positions that coincide well with the
characteristic activation energies of SmB$_6$ (Fig.~\ref{fig-pos}).
The important point here is that, whatever the interpretation of
these energies: the positions do not depend on the
zero-bias conductance, that means on the lateral contact size. It implies
that there is no additional voltage drop in the contact area, which could pose
a problem at very large junctions with $G_0 \ge 1\,$mS.
Thus Fig.~\ref{fig-pos} demonstrates that at least up to
$G_0 \approx 1\,$mS we can still derive the kinetic energy of the electrons
from the applied bias voltage because of the suppressed local heating.

Anomalies at $E_g^*$ as well as at $E_d^*$ could be observed at all junctions.
These dominant anomalies could indicate when the top of the valence band and
the narrow  impurity band of one electrode, respectively, face the bottom
of the conduction band of the other electrode. At some junctions we could
also resolve a small anomaly which corresponds to the lowest activation energy
$E_a$ of our sample. It  may represent an additional impurity band close
to the $E_d$ band.

Not all junctions showed anomalies at  around 0.2\,meV, 2\,meV, and 22\,meV.
An obvious reason for the sometime missing low-energy anomalies could be
the degradation of the interface region due to the contact, increasing the
local conductivity and thereby suppressing the low-energy processes.
Two of the anomalies, the one at 1.8\,meV and the other at 22\,meV,
have no counterpart as an activation energy from Fig.~\ref{fig-activate}.
And it is unclear whether they
just accidentally coincide with $E_g^*/2 - E_d^*$ and $2E_g^*$, respectively.
To speculate whether the anomaly at twice $E_g^*$ results from a double
junction, that is a contact with two junctions in series, can be safely
discarded because there are no corresponding anomalies at lower energies
at $2E_d^*$ or at $2E_a$.
The two above anomalies could be artefacts, created by the stress in the contact
region, and which is not present in the bulk sample. But there is no reason
why in such a case the position of the other anomalies should not be affected
in a similar way. This open question leads us to reconsider our way of
attributing certain energies derived form the activation energy to the
observed anomalies.

\begin{figure}
\centerline{\includegraphics[width=7cm]{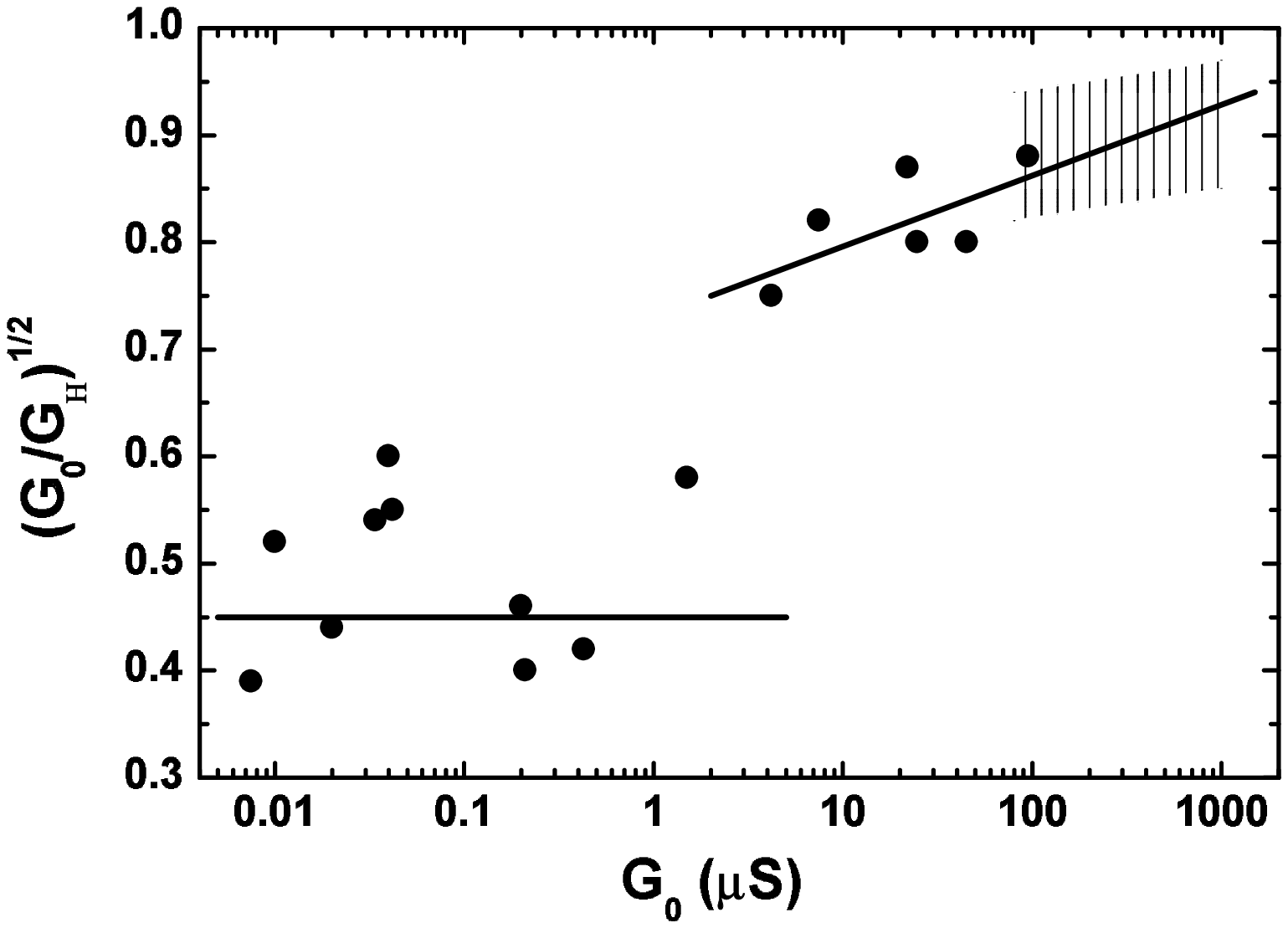}}
\caption{
Square-root of the normalized residual zero-bias conductance $\sqrt{G_0/G_H}$
vs.~zero-bias conductance $G_0$.
Without the gap a  zero-bias conductance $G_H$ is estimated.
Solid circles represent break-junction data, the shaded area indicates average
values derived from the spear-anvil type junctions (for which $G_0$ is
the zero-bias conductance at 4.2\,K). The solid lines are guides to the eye.}
\label{fig-dos}                              
\end{figure}

The necessary additional information can be obtained from the tunnelling
spectra.
As discussed above, a residual quasiparticle density of states at the Fermi
level of about $45\%$, derived from the size of the zero-bias anomlies of
small junctions with $G_0 \le 1\,\mu$S and shown in Fig.~\ref{fig-dos}, is
considerably smaller than the data obtained by conventional tunnel junctions
\cite{Amsler98}.
However, this difference is not unreasonable because the latter experiments,
compared to ours, were carried  out at much higher temperatures when the
$E_d$ anomaly has not fully developed.
The rather large scattering of our data points  in Fig.~\ref{fig-dos} probably
results, in a minor part, from our method of extrapolating $G_H$, the zero-bias
conductance without gap.
The major part is due to the fact that the break junctions are a local probe.
Each junction has slightly different local properties, for example due to
stress induced by the contact.
At planar tunnel junctions, on the other hand, the density of states is sampled
and averaged over a much wider contact area. Of course, uniform stress and
a preferred direction could affect the spectra of those planar  junctions as
well.

\begin{figure}
\centerline{\includegraphics[width=7cm]{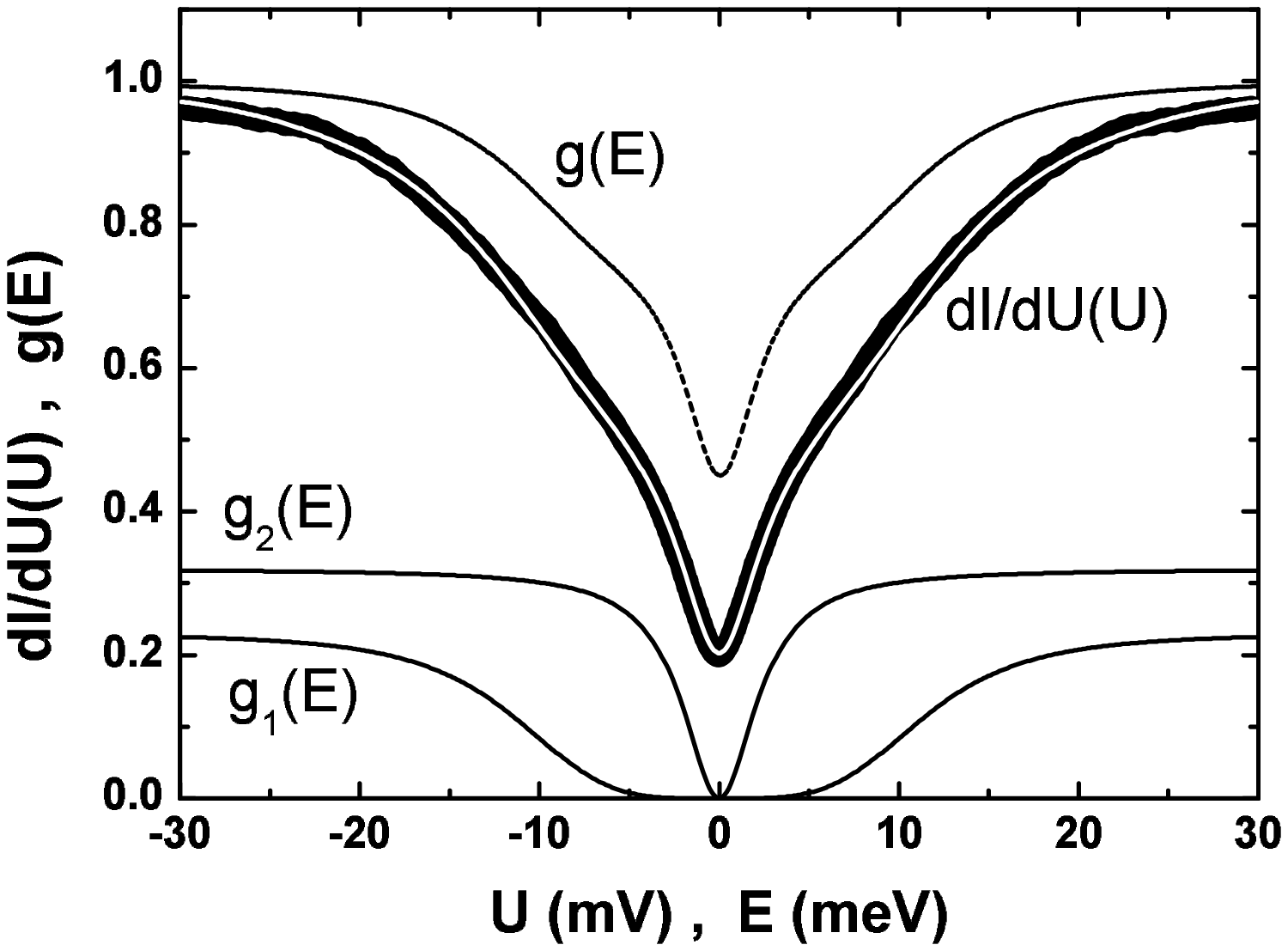}}
\caption{
Average $dI/dU$ tunnel spectrum (solid circles) and a fit (white solid line
through the data points) calculated using the density of states $g(E)$
shown by the dotted line. $g(E)$ is the sum of $g_1(E)$ and $g_2(E)$ plus
a constant background.}
\label{fig-fit}                              
\end{figure}

To estimate the average density of states at our SmB$_6$ tunnel junctions, that
means junctions with $G_0 \le 1\,\mu$S, we normalized their spectra with
respect to the voltage-dependent background as described in Fig.~\ref{fig-hi-G},
averaged the  normalized spectra, and then fitted the average
spectrum by various functional dependencies of the density of states,
assuming a constant transmission probability.
A good fit was obtained by assuming a constant background of
$45\%$ and two different  energy-dependent parts  $g_{1,2}(E)$ of the density
of states as shown in Fig.~\ref{fig-fit}.

The one part, $g_1(E)$, has a rather wide gap of about 21\,meV (full width at
half maximum), while the other part, $g_2(E)$, has a much narrower gap of
about 4.5\,meV. These two gaps may be attributed to the hybridization gap and
the $E_d$ anomaly, respectively. But in this case the width of the larger
anomaly differs a lot, almost a factor of two, from the activation energy data.
On the other hand, if the gap in $g_2(E)$ was suppressed, leaving only the
gap in $g_1(E)$, the anomaly of the density of states found in
Ref.~\cite{Amsler98} for planar SmB$_6$ - Pb tunnel junctions at 10\,K would
be reproduced quite well (the anomaly described by Ref.~\cite{Amsler98}
had a somewhat smaller width of about 14\,meV, possibly because it was also
asymmetric). This supports our interpretation of tunnelling due to the lateral
confinement at our junctions. It seems as if the gap in $g_2(E)$ develops only
at low temperatures $T \ll 10\,$K. And this  has to be expected  in view of the
bulk conductivity and the activation energy.

\begin{figure}
\centerline{\includegraphics[width=7cm]{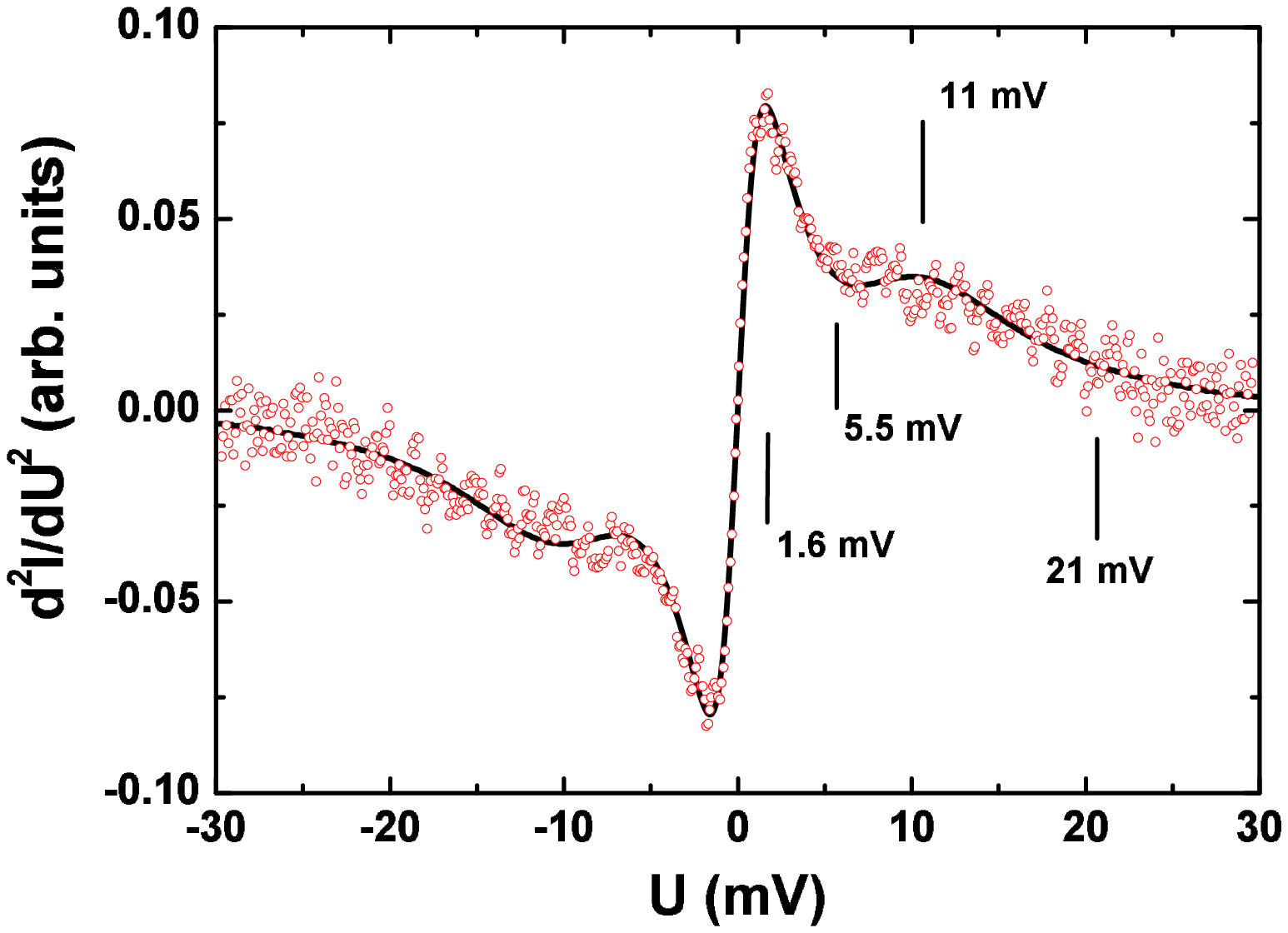}}
\caption{
Second derivative $d^2I/dU^2$ of the average spectrum of the tunnel junctions
(open sircles) and of the fit (solid line).}
\label{fig-2d}                              
\end{figure}

The second derivative $d^2I/dU^2$ of the average spectrum as well as of the
fit in Fig.~\ref{fig-2d} clearly show anomalies at around 1.6\,mV, 5.5\,mV,
11\,mV, and 21\,mV, respectively.
Thus except for $E_a = 0.2\,$eV  all anomalies found in the spectra as
displayed in Fig.~\ref{fig-pos} can also be recovered from the fit curve.
We believe that the activation-derived gaps differ from the real gap
values $E_g = 21\,$meV and $E_d = 4.5\,$meV because even at low temperatures
the density of states remains finite inside the gaps. Therefore electrons can
be thermally excited into states with energy $E \le E_d$ or $E \le E_g/2$,
respectively, smearing out the otherwise discrete activation-energy levels.

\section{Conclusion}
The differential conductance of SmB$_6$ - SmB$_6$ junctions has been
investigated at low temperatures as function of contact size.
A very wide range of data on the conductance scale down to very small junctions
was made possible because of the excellent mechanical stability of the break
junctions when compared to the spear-anvil type contacts.
Two different regimes of charge transport were distinguished.
Large junctions are in the diffusive Maxwell regime, in which the conductance
is dominated by the bulk conductivity. Local heating is negligible
because of the large phonon heat conductivity. Therefore, the applied bias
voltage can still be attributed to the kinetic energy of the charge carriers,
enabling spectroscopy.
Small junctions are in the tunnelling regime, although there is no dielectric
barrier. Depending on the model used, electrons tunnel through the contact
either by a  single hopping  event  like in the bulk material or as evanescent
waves because of  a large Fermi wavenumber.
Both models predict the transition to tunnelling when the zero-bias conductance
becomes smaller than about $10\,\mu$S, in good agreement with our experiments.
The spectra of these tunnel junctions indicate a finite $45\%$ residual
density of states at the Fermi level.
In the two different regimes of transport, at large as well as at small
junctions, anomalies can be resolved at the same energies.
However, only the spectra in the tunnelling regime  reveal the correct
gaps in the density of states of  $E_g = 21\,$meV and $E_d=4.5\,$meV that
are  responsible for the observed anomalies.
The absolute values of these two energies fit well those derived from other
experiments.

\section{Acknowledgments}
This work was supported by the Slovak Scientific Grant Agency VEGA, contract
1148 - 01 and 7022 - 20, and by the SFB 252 Darmstadt / Frankfurt / Mainz.
US Steel Ko\v{s}ice sponsored part of the liquid  nitrogen.



\end{document}